\documentclass[twoside,slac_one]{revtex4}
\usepackage{graphicx}
\usepackage{fancyhdr}
\usepackage{amsmath} 
\usepackage{bm}
\usepackage{amsxtra}
\usepackage{amssymb}
\usepackage{amsthm}
\usepackage{latexsym}
\usepackage{lscape}

\begin{document}

\title{Measurement of the Proton's Weak Charge at the Qweak Experiment}

%

\author{Jean-Francois Rajotte\footnote{On behalf of the Qweak collaboration}}
\affiliation{Laboratory for Nuclear Science,  Massachusetts Institute of Technology, Cambridge, MA, USA}

\begin{abstract}
The Qweak experiment at Jefferson Laboratory measures the parity violating
asymmetry of polarized electrons scattering from a proton target at very
low momentum transfer.  In the Standard Model, this asymmetry reveals
the proton's coupling to the neutral vector current, the weak charge.
This value, measured directly for the first time, will provide a precision
test of the Standard Model and will constrain the possibility of relevant
physics beyond the Standard Model.  The planned precision will probe certain classes of new physics at the ~2 TeV scale. In order to challenge the precise predictions, the asymmetry will be measured with a 2.5 percent accuracy. To achieve such a precision, great care has to be taken on many aspects of the experiment.  The very low momentum transfer reduces the hadronic effects to the asymmetry and must be determined to half of a percent accuracy.  Beam stability is controlled and monitored constantly and background events are carefully studied.
\end{abstract}

\maketitle

\thispagestyle{fancy}


\section{Introduction}
\begin{figure}[h]
\centering
\includegraphics[width=135mm]{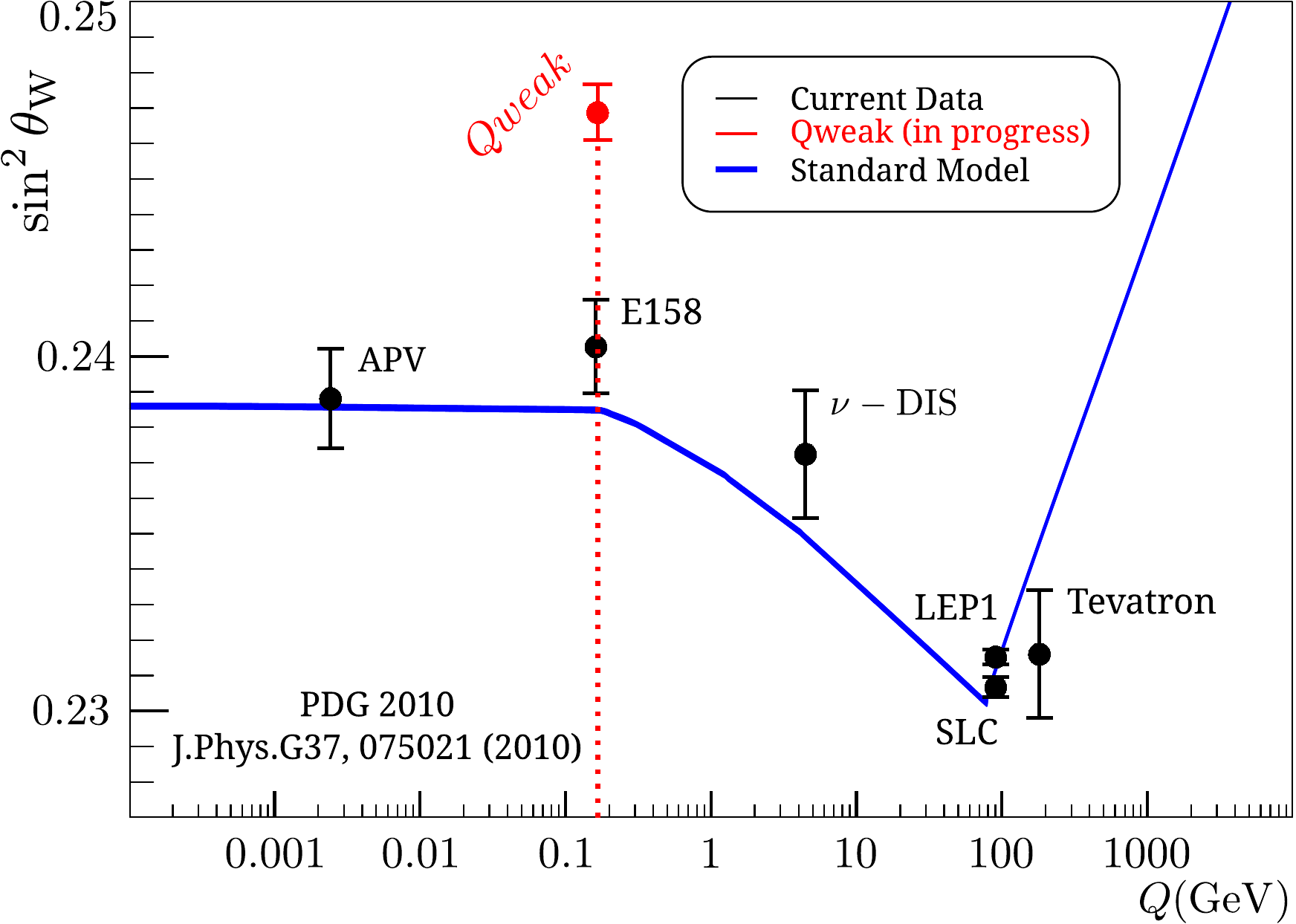}
\caption{Calculated running of the weak mixing angle in the Standard Model as function of the four-momentum transfer, $Q$, as defined in the modified minimal subtraction scheme \cite{Erler:2004in}.  The uncertainty in the predicted running corresponds to the thickness of the blue curve.  The black error bars show the current situation while the red error bar corresponds to the 4.1\% error expected (on $Q^p_W$) for the current measurement arbitrarily positioned on the vertical axis. } \label{sintheta_running}  
\end{figure}

The measurement of the proton's weak charge follows a long and rich tradition of parity violating electron scattering (PVES) experiments.  Homologous to the electric charge, the coupling strength of the electromagnetic interaction, the weak charge is the coupling strength of the weak interaction.  Table \ref{charge_table} shows the list of electromagnetic and weak charges for particles relevant to the present discussion\footnote{Only vector couplings are considered here since axial coupling is suppressed by the small vector charge of the electron.}.  The weak mixing angle, $\theta_W$, has a value such that $\sin^2\theta_W$ is about $1/4$, leading to a very small value of the proton's weak charge, $Q^p_{W}$.  This fortunate occurrence makes $Q^p_{W}$ very sensitive to $\sin^2\theta_W$.  The Standard Model predicts the running of $\sin^2\theta_W$ with very high precision as can be seen in Figure \ref{sintheta_running}.  A measurement of $\sin^2\theta_W$ deviating from this prediction would reveal the signature of physics beyond the Standard Model.  Conversely, a measurement in agreement with the Standard Model, in conjunction with existing measurements, will constrain the possibility of relevant new physics candidates.\\

\begin{table}[ht]
\begin{center}
\caption{Electroweak Couplings of $u$ and $d$ Quarks and Nucleons as Function of the Weak Mixing Angle $\theta_W$}
\begin{tabular}{|l|c|c|}
\hline \textbf{Particles} & \textbf{EM Charge} & \textbf{Weak Charge} \\
\hline $u$ & +$\frac{2}{3}$ & $1-\frac{8}{3}\sin^2\theta_W$ \\
\hline $d$ & $-\frac{1}{3}$ & $-1+\frac{4}{3}\sin^2\theta_W$ \\
\hline $p(uud)$ & $+1$ & $1-4\sin^2\theta_W$ \\
\hline $n(udd)$ & 0 & $-1$ \\
\hline
\end{tabular}
\label{charge_table}
\end{center}
\end{table}

\begin{figure}[h]
\centering
\includegraphics[width=80mm]{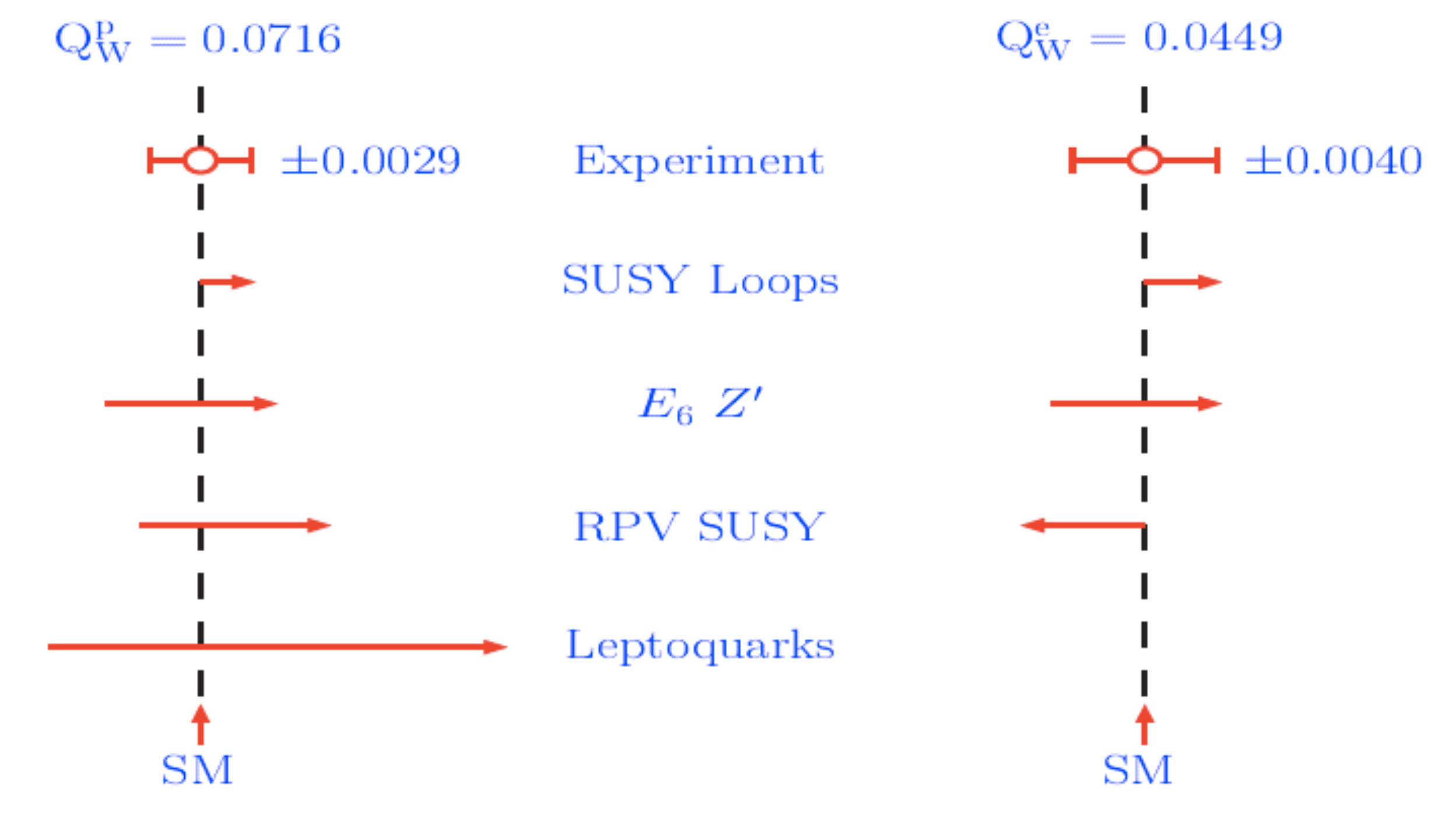}
\caption{Deviation from the Standard Model expected from various physics extensions for the anticipated 4.1\% error on the proton's weak charge and the most recent measurement on the electron's weak charge \cite{RamseyMusolf:2006vr}.  
Relative shifts induced by SUSY loops contributions are similar because the non-universal corrections experience significant cancellation.
Relaxing the assumption of R-parity conservation (RPV SUSY), allows further non-universal interactions that could affect $Q_W^p $ and $Q_W^e$ differently.  For the $Z^{\prime}$ theory based on the $E_6$ gauge group, the effect on both weak charges are also correlated, but can have either sign.  Leptoquark exchange through PVES leads to a significant contribution only to the weak charge of the proton because the measurement involves leptons and quarks at tree level.
} \label{QwPvsE}  
\end{figure}

There are several new of physics candidates that predict a deviation of $\sin^2\theta_W$ from the value calculated with the Standard Model.  Some examples are listed, along with their predicted variations based on the measured $Q^p_W$ for Qweak's expected error, in Figure \ref{QwPvsE}.  While the reader is referred to \cite{RamseyMusolf:2006vr} for the description of each mechanism, it is worth pointing out the complementarity of measuring the weak charge of the proton in parallel with the weak charge of the electron.  Indeed, their deviations from the Standard Model predictions are different for each mechanism.  For example, the proton's weak charge will be more sensitive to leptoquarks since the measurement contains leptons (the beam) and quarks (the target) at first order interaction whereas the electron's weak charge involves only leptons.  Another scenario of complementarity is the R-parity violating supersymmetry (RPV SUSY) that violates either the baryon or lepton quantum number.  In this type of supersymmetry, the proton and electron's weak charge difference from the SM prediction would be opposite in sign.  Thus, a deviation seen only on the proton's weak charge (the leptoquark scenario) or with opposite sign to the electron measurement (the RPV SUSY scenario) would be valuable information to discriminate processes of physics beyond the Standard Model.\\

\begin{figure}[h]
\centering
\includegraphics[width=80mm]{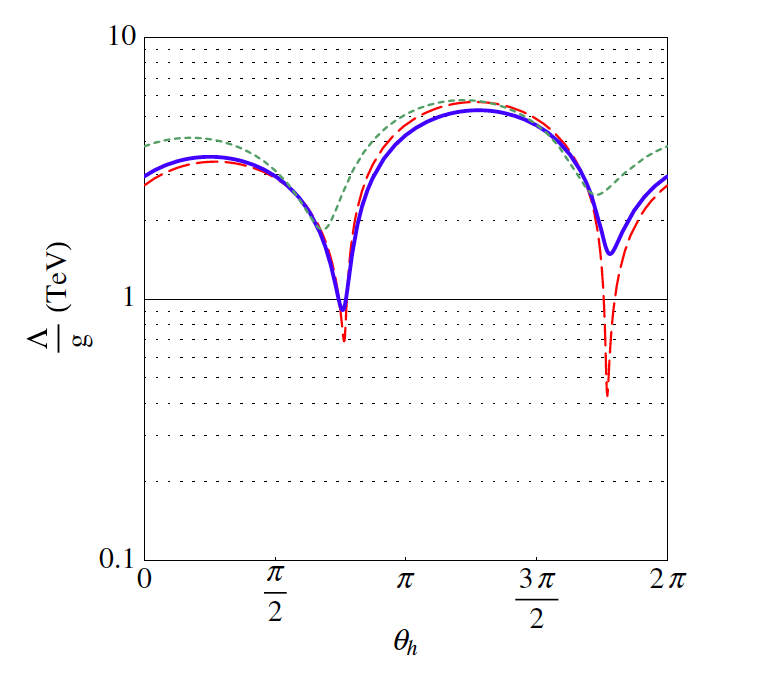}
\caption{Constraints on model independent parity-violating physics beyond the standard model in terms of mass scale versus the flavor mixing angle $\theta_h$ \cite{Young:2007zs}.  New parity-violating physics is ruled out at the 95\% confidence level below the curves.  The long-dashed red curves shows the current limit without parity-violating electron scattering data.  The solid blue curve shows the current limit including parity-violating electron scattering experiment.  The short-dashed green curve display the reach of the Qweak experiment assuming agreement with the Standard Model is observed.} \label{TeVlimit}  
\end{figure}

New physics can also be investigated through a model independent analysis.  Figure \ref{TeVlimit} shows how Qweak's measurement will be sensitive to the ratio of a new interaction scale, $\Lambda$, over its coupling strength, $g$, (as function of the interaction mixing angle) determined by \cite{Young:2007zs}.  This measurement could improve the current limit by as much as 1 TeV at the 95\% confidence level.\\

\section{Measuring the proton's weak charge through PVES}
The Qweak experiment measures the weak charge of the proton through parity violating of longitudinally electrons scattering from unpolarized protons in a liquid hydrogen target.  The setup has been designed to measure the rate of elastically scattered electrons.  
The measured rate asymmetry of alternating left and right helicity is related to the cross section asymmetry:
\begin{equation}
A_{LR} = \frac{\sigma_R - \sigma_L}{\sigma_R + \sigma_L}
\end{equation}
where $\sigma_L$ and $\sigma_R$ are the cross sections for right- and left-handed electrons, respectively.
For elastic scattering, the measured asymmetry can be written as follow:
\begin{equation}
A_{LR} = \left[\frac{-G_FQ^2}{4\sqrt{2}\pi\alpha}\right]
\frac{\varepsilon G^{\gamma}_EG^{Z}_E + \tau G^{\gamma}_MG^{Z}_M - (1-4\sin^2\theta_W)\varepsilon ^{\prime}G^{\gamma}_MG^{e}_A}
{\varepsilon(G^{\gamma}_E)^2 + \tau (G^{\gamma}_M)^2}
\end{equation}
where $G_F$ is the Fermi coupling constant, $Q^2$ is the four-momentum transfer squared and $\alpha$ is the fine structure constant.
The kinematic factors, functions of $Q^2$ and the scattering angle (in the laboratory frame), $\theta$, are defined as follow,
\begin{eqnarray}
&\tau  &=  \frac{Q^2}{4M^2}  \nonumber \\
&\varepsilon &= \frac{1}{1+2(1+\tau)\tan ^2\frac{\theta}{2}}   \nonumber \\
&\varepsilon ^{\prime} &= \sqrt{\tau(1+\tau)(1-\varepsilon ^2)}
\label{eq_AkinFactors}
\end{eqnarray}
The quantities $G^{\gamma}_E$, $G^{Z}_E$, $G^{\gamma}_M$, $G^{Z}_M$ are the vector form factors of the proton associated with $\gamma$- and $Z$-exchange and $G^e_A$ is the electron axial form factor.\\

\begin{figure}[h]
\centering
\includegraphics[width=80mm]{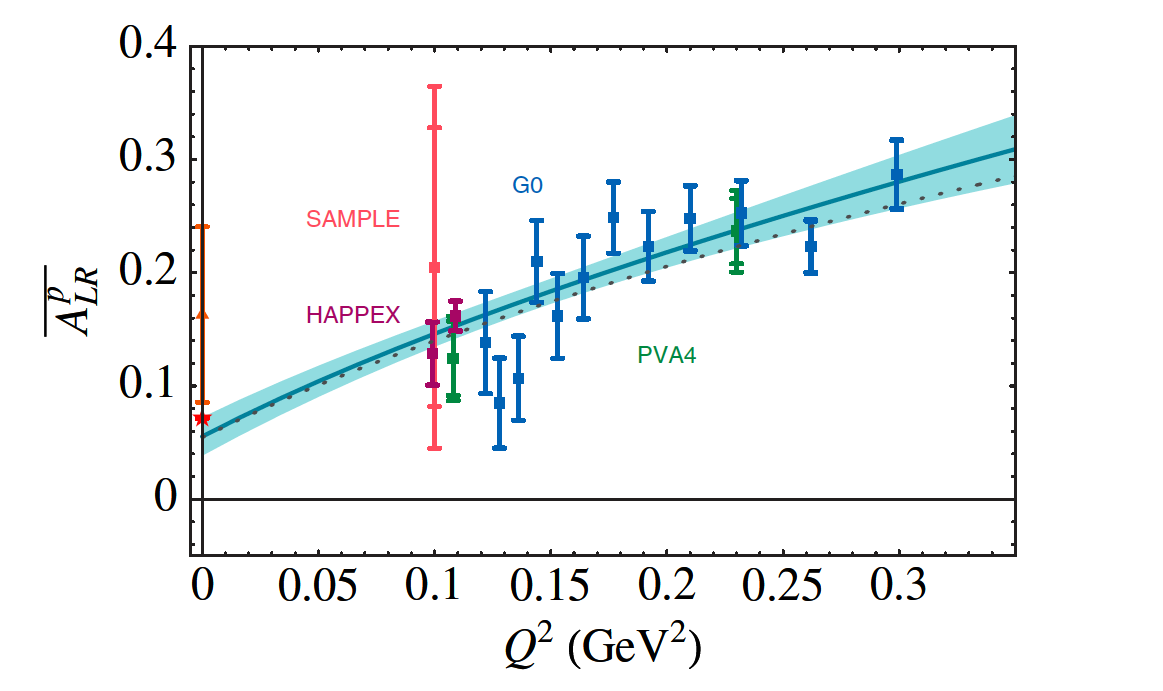}
\caption{Normalized parity violating asymmetry ($\overline{A^p_{LR}}\equiv A^p_{LR}/(A_0Q^2)$) measurements on a proton target, extrapolated to the forward-angle limit using world data on PVES as of 2007 \cite{Young:2007zs}.  The extrapolation to $Q^2=0$ measures the proton's weak charge.  The red star represents the Standard Model prediction and the orange solid line is the PDG Handbook value of the 2006 edition.  The solid blue line and its shaded error region represents the best fit on global electroweak data.  The dotted line close to the solid line is the same fit with an estimate of the anapole form factors of the proton.} \label{q2_extrapol}  
\end{figure}

In order to select elastically scattered electrons, the energy of the incident beam has been chosen to be 1.165\,GeV and the scattering angle to be about 8 degrees leading to a small $Q^2$ of about 0.026 (GeV/c)$^2$.  In this kinematic range, the measured asymmetry can be rewritten explicitly as function of the proton's weak charge \cite{Young:2007zs}:
\begin{equation}
A_{LR} \simeq A_0\left[Q^p_{W}Q^2 + BQ^4+\ldots \right]
\end{equation}
where $A_0= -G_F/(4\pi\alpha\sqrt{2})$ and the second term, $B$, contains the effect of the hadronic structure of the proton.  The proton's weak charge is determined by extrapolating the measured asymmetry to $Q^2=0$.  This is done with the form factor measurements by previous PVES experiments as shown in Figure \ref{q2_extrapol}.\\

\section{Experiment}
The Qweak experiment, conducted in Hall C at Jefferson Laboratory, can be reduced to three components: 
a longitudinally polarized electron beam, a liquid hydrogen target and a detector system.  Each of these components needs to meet its specification to reach the stringent goal of measuring the parity violating asymmetry of about 230 part-per-billion (ppb) with a 2.5\% precision.  
The detailed error budget expected for the asymmetry determination and its translation into $Q^p_W$ uncertainties is shown in Table \ref{t_error_budget}.\\

\begin{table}[ht]
\begin{center}
\caption{Detailed Error Budget Expected for the Qweak Experiment}
\begin{tabular}{|l|c|c|c|}
\hline  \textbf{Source of error} & $\Delta A_{LR}/A_{LR}$  & $\Delta Q^p_{W}/Q^p_{W}$ \\
\hline  
Counting Statistics & 2.1\%  & 3.2\% \\
Hadronic Structure & -  & 1.5\% \\
Beam polarimetry & 1.0\%  & 1.5\% \\
Absolute $Q^2$ & 0.5\%  & 1.0\% \\
Backgrounds & 0.5\%  & 0.7\% \\
Helicity-correlated beam properties & 0.5\%  & 0.7\% \\
\hline  \textbf{TOTAL} &  \textbf{2.5\%}  &  \textbf{4.1\%} \\
\hline
\end{tabular}
\label{t_error_budget}
\end{center}
\end{table}

The polarized electron source is based on photoemission by polarized light from a GaAs crystal.  The high statistical needs of Qweak are fulfilled by a high current, between 150 and 180\,$\mu$A, passing through the target.  The helicity of the electron beam is controlled by the polarity of the voltage applied to a Pockels cell in the laser beam changing at a rate of 960\,Hz.  A set of monitors in the beam line measures the characteristics of the beam such as intensity and position which are also used to extrapolate the beam properties at the target.  A position monitor located at a point where the beam is dispersed in momentum serves to analyze the energy.\\

The beam hits a 35\,cm long aluminum cell filled with liquid hydrogen (LH$_2$).  The LH$_2$ is maintained at 20K by a cryogenic system of 2500W cooling power.  The target cell has been designed such that inhomogeneities in pressure and temperature are minimized.  The variations in the density of the target induced by the beam are slower than the helicity reversal, making the target stable during an asymmetry measurement (performed over four helicity windows).  In order to keep hydrogen from freezing, a heater takes over at the moment the beam switches off.\\

\begin{figure}[h]
\centering
\includegraphics[width=135mm]{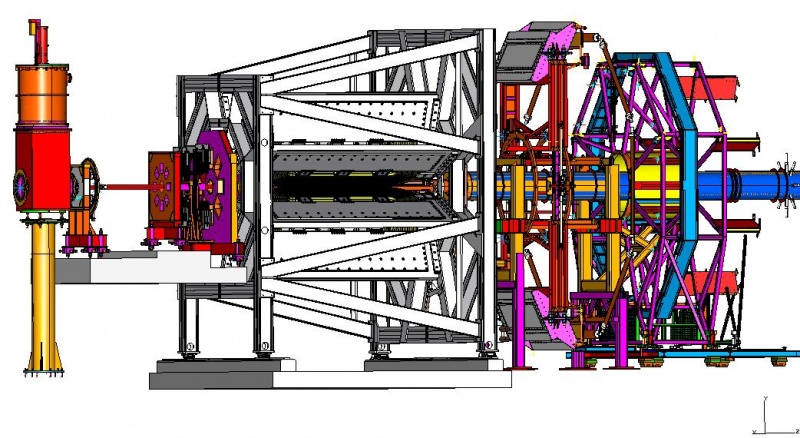}
\caption{Layout of the Qweak experiment without the shielding.  The electrons travel from left to right through the target, the first collimator set and the drift chambers, the toroidal magnet, the downstream drift chambers and the quartz \v{C}erenkov detector in dark red (not all elements are mentioned).  The drift chambers are used only at lower current (the so-called counting mode) to determine the $Q^2$.} \label{qweak_exp}  
\end{figure}

The scattered electrons exit the target at an angle of about 8 degrees to be measured by the detector system represented in Figure \ref{qweak_exp}.  The elastically scattered electrons are selected by a series of collimators which define the $Q^2$ acceptance.  An 8 segment toroidal magnet focuses the elastic electrons through another set of collimators onto the 8 main \v{C}erenkov detectors.  Each of the \v{C}erenkov radiator bars is $200\times18\times1.25$\,cm$^3$ and yield about 16 photo-electrons per track.  At the end of each bar, a 5\,inch photo-multiplier tube collects the photo-electrons.  These bars work in an integrating mode, i.e.\ individual events are not counted but, rather, the integrated response over the beam burst is recorded.  In order to increase the electron signal and reduce pion background, a 2\,cm thick Pb pre-radiator was installed in front of each radiator bar.  The pre-radiator, combined with further shielding, reduced soft background from about 10\% to about 0.1\%.\\

In an ideal experiment, the above mentioned components would be enough, but backgrounds and variation of many parameters such as beam polarization and helicity correlated beam properties impose further considerations.  Three important features of the experiment must be known at a precision level that is coherent with the expected error of the asymmetry shown in Table \ref{t_error_budget}.  The measured asymmetry, $A_{meas}$, is related to the PVES asymmetry as follow:
\begin{equation}
A_{meas} = P(1-f)A_{LR}(Q^2) + fA_{bkg} + A_{false}
\label{eq_ameas}
\end{equation}
where $P$ is the beam polarization and $f$ is the dilution factor (i.e.\ the fraction of background over signal+background).
Clearly, it is also essential to have an accurate knowledge of the polarization of the beam.  Qweak aims for a 1\% determination of the (often exceeded) 85\% polarization.  The first method of polarization measurement is through the M\o ller polarimeter, a well established method already used during previous experiments in Hall C.  The method consists of inserting a polarized iron foil along the beam trajectory and use the polarization dependent feature of the M\o ller cross section to deduce the beam polarization.  This method is invasive and must be performed at lower current (few $\mu$A) for about four hours.  With this technique, beam polarization is measured about three times a week during normal conditions and after any intervention in the beam line that could have affected the beam properties.  In order to determine the polarization while taking data, the Hall C beam line was equipped with a Compton polarimeter.  A circularly polarized laser beam hits the electron beam and the polarization dependent feature of the cross section is used to determine the beam polarization.  There are two detectors, one for the scattered electrons and one for the backscattered photons.  Both provide measurements of the polarization.  Since this can be done in nominal conditions, the Compton polarimeter can fill the gap between the precise M\o ller measurements.

The aluminum windows of the target cell are a major source of background because they contain neutrons which have a much stronger weak charge as can be seen in Table \ref{charge_table}.  
The effect of aluminum, leading to a correction of about 20\%, must be known at a precision of a few percent and is investigated through simulation and by measurement of an empty target cell.  Another element of interest in equation (\ref{eq_ameas}) is the $Q^2$ which has to be known at 1\% accuracy.  
The $Q^2$ determination is achieved with two sets of drift chambers, upstream and downstream of the toroidal magnet.  
Due to the high rate of nominal beam current ($150-180$\,$\mu$A), the measurement is performed at reduced current of about 50\, pA.  
There is also a quartz scanner of $1\times1\times2$\,cm$^3$ fused silica located behind the main detectors which can either run with the reduced current and the full 180\,$\mu$A.  Hence, the scanner asserts that the scattered electrons measured by the drift chambers have the same kinematic profile as in nominal data taking.

The last term of  equation (\ref{eq_ameas}) refers to correlations with helicity of various parameters such as beam energy, position and intensity.  It should contribute no more than 0.5\% of the final asymmetry measurement uncertainty.  These errors are controlled by, first, monitoring and minimizing them during data taking and by correcting the remaining asymmetries at the analysis stage.  In the analysis, the raw asymmetries are corrected using the equation:
\begin{equation}
A_{meas} = A_{raw} - \sum_i(a_i \delta M_i)
\label{eq_ameas_from_raw}
\end{equation}
where $A_{raw}$ is the uncorrected asymmetry, $\delta M_i$ are the differences in beam monitors correlated with helicity, and $a_i$ are the correction coefficients which are a measure of the sensitivity of the asymmetry to fluctuations in the beam parameters.  Since the $a_i$ are measured concomitantly with data taking, they are valid for exact running conditions.  
A further protection against helicity correlated properties relies on the reversal of the helicity of the beam by independent methods.  Using a half-wave plate at the source, the direction of the linear polarization of the laser light incident to the Pockels cell is inverted every eight hours.  Furthermore, an even slower reversal through Wien rotation along the beam line is applied every week.\\

\section{Results and Outlook}
\begin{figure}[h]
\centering
\includegraphics[width=135mm]{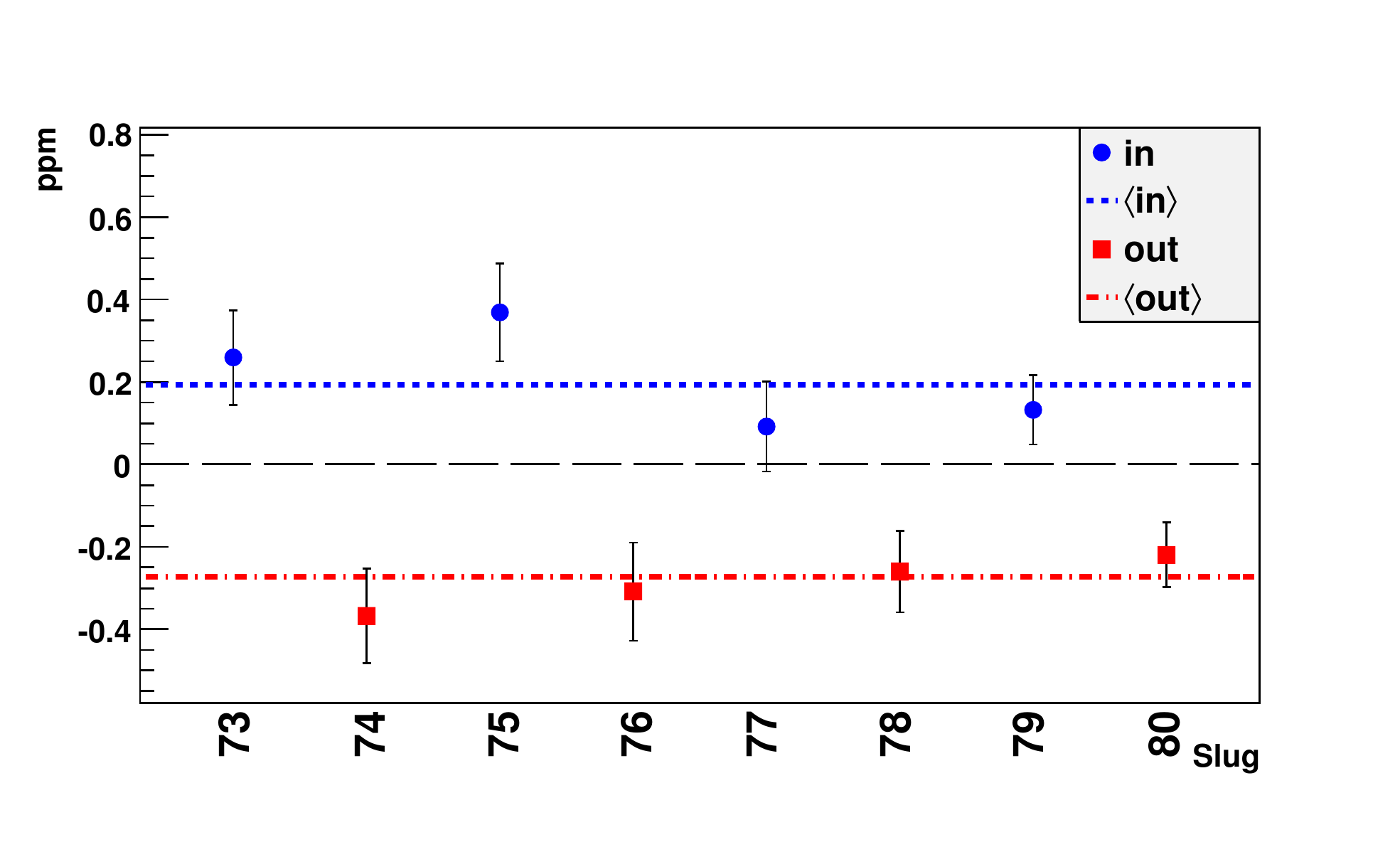}
\caption{Experimental raw asymmetries (cf. equation (\ref{eq_ameas_from_raw})) on all of the radiator bars as function of slug numbers which represents about 8 hours of data taking.  The position of the insertable half-wave plate (HWP), inverting the polarization state at the source, 
is indicated by the blue disk, HWP \textit{in}, and red square, HWP \textit{out}.  
One can clearly observe the sign flip of the asymmetries when the HWP is inserted.  
The dashed blue line shows the average of HWP \textit{in} asymmetries 
and the dashed-dotted red line the average of the HWP \textit{out} asymmetries.  
A blinding factor has been applied to the data.}
 \label{mdallbars_perSlugs}  
\end{figure}

The first phase of the experiment has finished in May 2011 and Qweak has accumulated about 25\% of the total statistical sample needed to achieve the 4\,ppb error on the asymmetry.  A selected sample of the measured asymmetries is shown in Figure \ref{mdallbars_perSlugs}, where the effect of the slow reversal of the beam helicity by the insertable half-wave plate can be seen.  
At the current of 165\,$\mu$A, the width of the measured asymmetry distribution is about 236\,ppm.  
Combining the width of pure counting statistics, 215\,ppm, to our determination of detector resolution, target fluctuations and current normalization, the width is expected to be 235\,ppm.  One therefore concludes that the resolution of the measurement is well understood.

The second and last phase will begin in November 2011 and finish in May 2012.  The required conditions of the experiment are met and Qweak is on its way for a 4\% measurement of the weak charge of the proton to challenge the Standard Model prediction.

\begin{acknowledgments}
I am grateful to my colleagues at Qweak who have assisted me with patience as I joined the experiment in a late stage.
\end{acknowledgments}

\bigskip 

\end{document}